\documentclass[twocolumn,preprintnumbers,amsmath,amssymb,superscriptaddress]{revtex4}
\usepackage{epsfig,bm}
\usepackage{ulem}
\usepackage{amsmath,color}

\begin{document}
\title{Low-lying electric-dipole strengths of Ca, Ni, and Sn isotopes
  imprinted on total reaction cross sections}
\author{W. Horiuchi}
\affiliation{Department of Physics,
  Hokkaido University, Sapporo 060-0810, Japan}
\author{S. Hatakeyama}
\affiliation{Department of Physics,
  Hokkaido University, Sapporo 060-0810, Japan}
\author{S. Ebata}
\affiliation{Nuclear Reaction Data Centre, Faculty of Science, 
  Hokkaido University, Sapporo 060-0810, Japan}
\author{Y. Suzuki}
\affiliation{Department of Physics, Niigata University, Niigata 950-2181, Japan}
\affiliation{RIKEN Nishina Center, Wako 351-0198, Japan}

\begin{abstract}
Low-lying electric-dipole ($E1$) strength of a neutron-rich nucleus 
contains information on 
neutron-skin thickness, deformation, and shell evolution.
We discuss the possibility of making use of total reaction cross sections 
on $^{40}$Ca, $^{120}$Sn, and $^{208}$Pb targets 
to probe the $E1$ strength of neutron-rich Ca, Ni, and Sn isotopes.
They exhibit large enhancement of the $E1$ strength at neutron number
$N>$ 28, 50, and 82, respectively, due to a change of the single-particle
orbits near the Fermi surface participating in the transitions.
The density distributions and the electric-multipole strength functions 
of those isotopes are calculated
by the Hartree-Fock$+$BCS and
the canonical-basis-time-dependent-Hartree-Fock-Bogoliubov
methods, respectively,
using three kinds of Skyrme-type effective interaction.
The nuclear and Coulomb breakup processes are respectively described
with the Glauber model and the equivalent photon method in which 
the effect of finite-charge distribution is taken into account.
The three Skyrme interactions give different results
for the total reaction cross sections because of different
Coulomb breakup contributions.
The contribution of the low-lying $E1$ strength 
is amplified when the low-incident energy is chosen.
With an appropriate choice of the incident energy and target nucleus,
the total reaction cross section can be complementary to the Coulomb 
excitation for analysing the low-lying $E1$ strength of unstable nuclei.
\end{abstract}
\maketitle

\section{Introduction}

Electric-dipole ($E1$) excitations of nuclei provide 
useful information on 
not only the ground-state properties but also
the dipole excitation mechanism.
In neutron-rich nuclei, the low-lying $E1$ strength, so-called 
pygmy dipole resonance, has attracted much attention.
In particular, the correlation between the low-lying $E1$ mode
and neutron-skin thickness has intensively been discussed
in relation to the equation of state of
asymmetric neutron matter~\cite{Reinhard10,Inakura13,RocaMaza15}.

The $E1$ excitation mechanism in the neutron-rich unstable nuclei
has not yet reached a universal understanding.
Recent systematic analyses of the $E1$ strength
show that its enhancement 
strongly depends on the shell structure
and mass region~\cite{Ebata14,Papa14,Papa15}.
A relationship with the so-called ``soft'' dipole mode due to 
the excess neutrons
and a core nucleus~\cite{Yamazaki86,Ikeda88, Suzuki90}
is also an interesting subject as a characteristic
excitation mode in the neutron-rich unstable nuclei
(See recent papers~\cite{Mikami14, Inakura14} and references therein).

Experimental studies of the low-lying $E1$ strength have been performed 
by using both photoabsorption reactions with real photons 
and Coulomb breakup reactions with virtual photons generated 
by a highly-charged target nucleus. 
The former cannot be applied to short-lived unstable nuclei,
while the latter can be applied in the inverse kinematics
and has often been utilized to extract
the low-lying $E1$ strength of halo nuclei~\cite{Nakamura06,Nakamura09}.
Since a weakly-bound halo nucleus breaks 
mainly through the $E1$ transition,
the $E1$ strength function is extracted by
subtracting the nuclear contribution 
from one- or two-neutron removal cross sections.
We can not apply this idea for our purpose because 
most of unstable nuclei are more tightly bound than
the halo nuclei. Higher multipole excitations
other than $E1$ can also be expected to play a significant role.
Furthermore, for the calculation of the nucleon-removal cross sections,
we have to assume appropriate final-state wave functions, 
leading to some ambiguity. See, e.g., Refs.~\cite{Ibrahim04,Capel08}.

Inclusive observables that require no final-state wave functions  
are desired to probe the low-lying $E1$ strength.
The total reaction or interaction cross sections measured at 
medium- to high-incident energies at $\sim$100-1000$A$\,MeV
are possible candidates for that 
purpose. They have been so far used primarily to study 
nuclear sizes thanks to the following advantages:
The measurement is easier and applicable to almost all nuclei, 
and theoretical models to evaluate the cross sections 
are well established. The cross sections for stable
and unstable nuclei have been measured
using light targets, e.g., $^{1}$H and $^{12}$C, because 
the Coulomb breakup contribution can be negligible.
The  measurement reaches few \% accuracy 
for unstable nuclei~\cite{Takechi10,Kanungo11,Kanungo11b,Takechi14}, 
and thus we can discuss structure problems including,  
for example, deformation~\cite{Minomo11,Horiuchi12,Minomo12,Sumi12,Watanabe14}
and neutron-skin thickness~\cite{Horiuchi14,Horiuchi16}.

In the above measurement of the reaction and interaction cross sections the 
target nucleus is chosen to be light enough to enable one to neglect 
the Coulomb breakup contribution~\cite{Horiuchi16}.
In this paper, we take an opposite direction: 
We instead consider a heavy target in which the reaction
includes large Coulomb breakup contributions,
and discuss the possibility of extracting
the low-lying $E1$ strength 
using the total reaction cross section. This is challenging in that 
the Coulomb breakup contribution has to be evaluated by a sound theory. 

The Coulomb breakup process in the high-energy collision 
is well approximated with
the equivalent-photon method (EPM)~\cite{Bertulani88,Suzuki03,Bertulani04}.
It is based on a semi-classical picture in which 
a relativistic charged particle passes through
the Coulomb field produced by the highly-charged target nucleus.
If the Coulomb excitation is dominated by the $E1$ process,
the Coulomb breakup cross section is simply obtained
by multiplying the number of
the virtual photons and the $E1$ strength functions.
The energy dependence of the number of the virtual photons is
important: At the high-incident energy $\sim$1000\,MeV/nucleon,
the number of the virtual photons distributes 
from the low-lying to the giant resonance energy region.
With the decrease of the incident energy,
the virtual photons are concentrated at the low energy region
and therefore the contribution of the low-lying strength function
to the Coulomb breakup cross sections will be more enhanced.
The EPM has been employed to extract the structure of halo nuclei
 from the low-lying $E1$ strength~\cite{Nakamura06,Nakamura09}.
Reasonable agreement between theory and recent experiment
is obtained~\cite{Nakamura09,Horiuchi10}
and its validity is studied by reaction
calculations~\cite{Yahiro11,Yoshida14}.
We will make use of this sensitivity for extracting the low-lying
$E1$ strength of unstable nuclei.

In this paper, we take up Ca, Ni, and Sn isotopes
with even neutron numbers $N=20$-40 for Ca, 28-56 for Ni, and 50-90 for Sn,
because the enhancement of the low-lying $E1$ strength is predicted 
at $N>28$, 50, 82 for Ca, Ni, and Sn isotopes, respectively~\cite{Ebata14}.
At those magic numbers, $0f_{7/2}$, $0g_{9/2}$, and $0h_{11/2}$
neutron orbits are fully occupied, respectively.
The higher major shell orbits play
a vital role to determine the $E1$ strength.
A sudden increase of the $E1$ strength is clearly seen
in ``PDR fraction''~\cite{Ebata14},
which is defined as a fraction of
total and cumulative energy-weighted
sums up to {\it ad hoc} cut-off excitation energy, 10\,MeV. 
The enhancement in fact strongly depends on the interaction employed
or very sensitive to the single-particle structure
near the Fermi surface.
We investigate the total reaction cross sections including
the Coulomb multipole excitations in order to answer whether
they can be used as a probe of the low-lying $E1$ strength.
This study provides us with information on the shell structure beyond
  the magic numbers $N=28$, 50, 82 of neutron-rich nuclei and
  can also be a strong test of the Skyrme interaction employed.
The role of the low-lying $E1$ strength is quantified
in the Coulomb breakup contribution as well as the
contributions from other electric multipoles.
Incident energy and target dependence of
the total reaction cross sections
are systematically analysed.

In the next section, we describe our reaction and structure models 
as well as a way to include the Coulomb multipole effects
into the cross section.
Since we consider highly-charged particles,
a finite size effect of the target charge distribution, which
is usually ignored in the EPM, is also formulated in this section.
Numerical results are presented in Sec.~\ref{results.sec},
mainly focusing on Sn isotopes.
In Sec.~\ref{finite.charge.dist},
we first discuss the effect of the finite charge distribution
in the EPM.
In Sec.~\ref{rcs.sec}, we make a systematical analysis of 
the total reaction cross sections of Sn isotopes.
The contributions of the Coulomb multipole excitations
are quantified in Sec.~\ref{multi.sec}.
We discuss in Sec.~\ref{resp.sec} the incident energy and target dependence 
of the total reaction cross section and its sensitivity to the $E1$ strength.
In Sec.~\ref{rcsCaNi.sec},
we show the total reaction cross sections of Ca and Ni isotopes.
Conclusions are given in Sec.~\ref{conclusions.sec}.

\section{Total reaction cross section}
\label{methods.sec}

We consider the total reaction cross section ($\sigma_R$) on 
a heavy target that induces a large amount of Coulomb excitations.
The total reaction cross section 
is expressed as a sum of the nuclear breakup cross section ($\sigma_N$) and 
the Coulomb breakup cross section ($\sigma_C$): 
\begin{align}
  \sigma_{R}=\sigma_N+\sigma_C.
\end{align}
See, e.g., Refs.~\cite{Ibrahim00,Horiuchi16} for its validity.
The nuclear and Coulomb 
interference term is negligibly small. 
These cross sections are calculated as explained below. 

\subsection{Nuclear breakup}

The $\sigma_N$ is calculated in 
the Glauber formalism~\cite{Glauber} by
\begin{align}
\sigma_{N}=\int d\bm{b}\, (1-|e^{i\chi(\bm{b})}|^2),
\end{align}
where $\bm{b}$ is the impact parameter vector
perpendicular to the beam direction.
The nuclear optical phase-shift function, $\chi(\bm{b})$,
contains all information of the high-energy nuclear collision.
We calculate $\chi(\bm{b})$ in 
the Nucleon-Target formalism in the Glauber theory~\cite{NTG}, which 
is known to give a better description of high-energy nucleus-nucleus
collisions than the ordinary Optical-Limit-Approximation. It is 
easily calculated by using the ground state densities of both the projectile and target nuclei and the
parameters of the profile function describing the $NN$ collision 
are taken from Ref.~\cite{Ibrahim08}. 
The present method for computing $\sigma_N$ 
has been successfully applied to many examples
of nucleus-nucleus collisions including
light unstable nuclei~\cite{Horiuchi06,Horiuchi07,Ibrahim09,Horiuchi10,Horiuchi12}.

We use the ground-state density distributions of Ca, Ni, and Sn isotopes
and the target nuclei, $^{40}$Ca, $^{120}$Sn, and $^{208}$Pb
obtained in Ref.~\cite{Horiuchi16}, where  
the Hartree-Fock (HF)$+$BCS method is applied to 
three kinds of the Skyrme-type effective interaction,
SkM*~\cite{SkMs}, SLy4~\cite{SLy4}, and SkI3~\cite{SkI3}.
We employ a constant monopole pairing
  as in Refs.~\cite{Ebata14,Ebata17}, where the level density
  determining its pairing strength is calculated
  by each of the Skyrme interactions.
Once all the inputs are set, the calculation of $\sigma_N$ contains 
no adjustable parameter.

\subsection{Coulomb breakup}

To calculate $\sigma_C$ we have to consider some basic elements such as 
equivalent photon method (EPM), 
photoabsorption cross sections, and effect of finite charge 
distribution. These are discussed below. 

\subsubsection{Multipole excitations by virtual photons}

We consider the Coulomb breakup probability 
$P_C(\bm b)$ according to the EPM~\cite{Bertulani88,Suzuki03,Bertulani04}.
The Coulomb breakup occurs through 
both electric- and magnetic-multipole excitations, but the latter contribution 
is ignored in this paper 
because a ratio of the photon-number spectra
  of $E1$ and $M1$ transitions
  is roughly proportional to $(v/c)^4$~\cite{Bertulani88},
  and the $M1$ strength is in general
  much smaller than the $E1$ strength~\cite{BM}.  
$P_C(\bm b)$ is given as a sum of electric multipoles labeled by $\lambda$,
and each multipole is obtained by
the equivalent photon number $N_{E \lambda}(\bm{b},\omega)$ multiplied by
the photoabsorption cross section $\sigma_{E \lambda}(\omega)$
integrating over the frequency $\omega$: 
\begin{align}
 P_C(\bm{b})
 =\sum_{\lambda}
 \int_0^\infty d\omega\,  N_{E \lambda}(\bm{b},\omega)\sigma_{E \lambda}(\omega).
 \label{EPMprob.eq}
\end{align}
Assuming point-charge distribution
    of the target nucleus, the multipole decomposition
of the photon numbers per unit area per unit frequency
is given by~\cite{Bertulani88} 
\begin{align}
 &N_{E\lambda}(\bm{b},\omega) \notag\\
  &=Z_T^2 \alpha \frac{\lambda[(2\lambda + 1)!!]^2}{(2\pi)^3 (\lambda + 1)}
  \sum_m |G_{E \lambda m}(\xi)|^2 \frac{\xi^2}{\omega b^2} K_m^2 \left(\xi \right)
\label{photon.eq}
\end{align}
with
\begin{align}
  &G_{E\lambda m}(x)=i^{\lambda+m} \frac{\sqrt{16\pi}}{\lambda(2\lambda+1)!!}\notag\\
&\times \left\{\frac{(\lambda+1)(\lambda+m)}{2\lambda+1}P_{\lambda-1}^{m}(x)-\frac{\lambda(\lambda-m+1)}{2\lambda+1}P_{\lambda+1}^{m}(x) \right\},
\end{align}
where $\alpha$ is the fine structure constant and $\xi=b\omega/\gamma v$ with  
the Lorentz factor $\gamma=1/\sqrt{1-(v/c)^2}$. $K_m$ is the modified Bessel function of the second kind and $P_{l}^{m}$ is the associated Legendre polynomial. 

\subsubsection{Mean-field calculations for photoabsorption cross sections}

The nuclear structure information of the Coulomb breakup reaction is 
contained in $\sigma_{E \lambda}(\omega)$, which is related to the $E\lambda$ strength (response) function $S(E\lambda; \omega)$ as
\begin{equation}
\sigma_{E \lambda}(\omega)= \frac{(2\pi)^3 (\lambda+1)}{\lambda[(2\lambda+1)!!]^2}\omega^{2\lambda-1} S(E\lambda; \omega).
\end{equation}
The canonical-basis-time-dependent-HF-Bogoliubov 
method is employed to obtain $S(E\lambda; \omega)$~\cite{Ebata10}.
A linear response by the $E\lambda$ field is obtained
using the prescription given in Ref.~\cite{NY05}.
The initial state is generated by
applying a weak impulse field to the ground state: 
\begin{align}
  F^{E1}_K&=
  \begin{cases}
    e(N/A)rY_{1K}(\hat{\bm{r}}),\quad ({\rm for\ proton})\\
    -e(Z/A)rY_{1K}(\hat{\bm{r}}),\quad ({\rm for\ neutron})
  \end{cases}\\
  F^{E\lambda}_K&=e\frac{r^\lambda Y_{\lambda K}(\hat{\bm{r}})
    +r^\lambda Y_{\lambda -K}(\hat{\bm{r}})}{\sqrt{2(1+\delta_{K0})}},
  \quad ({\rm for\ proton},\  \lambda > 1)
\end{align}
and the time evolution of the initial state enables us to obtain 
the strength function.

\subsubsection{Equivalent photon method with finite-charge distribution}

If the target nucleus is treated as a point-charged particle
with charge $Z_Te$, 
the number of equivalent photons at the center-of-mass ($\bm r=0$)
of the fast-moving projectile nucleus with velocity $v$
is obtained by using the electric field ${\bm E}(\bm r, \omega)$ as~\cite{Bertulani88,Suzuki03}  
\begin{align}
N(\bm b, \omega)&=\frac{c}{\hbar \omega}
|{\bm E}(\bm r, \omega)|^2_{\bm r=0} \notag \\
&=\frac{Z_T^2\alpha}{\pi^2}\big(\frac{c}{v}\big)^2 \frac{\xi^2}{\omega b^2}\left[
K_1^2(\xi)+\frac{1}{\gamma^2}K_0^2(\xi)\right].
\label{EPMpoint.eq}
\end{align}
As was done in Ref.~\cite{Bertulani88},
the multipole decomposition of the electric field is possible
by considering $\bm{r}$-dependence but
we discuss the electric field at the origin $\bm{r}=0$ in
this paper for the sake of simplicity. Note that
$N_{E1}(\bm{b}, \omega)$ of Eq. (\ref{photon.eq})
is equal to $N(\bm b, \omega)$.

The target nuclei considered in this paper are medium and heavy nuclei, and 
it is appropriate to discuss possible deviation 
from the point-charge approximation. In the following 
we estimate the extent to which 
$N(\bm b, \omega)$ of Eq.~(\ref{EPMpoint.eq}) changes for 
the finite charge distribution. 
Let $Z_Te\rho_T(\bm r')$ denote the charge density of the target 
nucleus, $\int d{\bm r}' \rho_T(\bm r')=1$. The electric field 
produced by the fast moving target nucleus is 
\begin{align}
\bm E(\bm r, t)|_{\bm r=0}=-Z_Te\int d{\bm r}' \frac{\bm R(t)}{\gamma^2 u^3}
\rho_T(\bm r'),
\label{efzero}
\end{align}
where, with $\bm r'=(\bm s', z')$,  
\begin{align}
&\bm R(t)=\bm b + \bm v t+\bm s'-\frac{\bm v}{v}z', \notag \\
&u=\sqrt{\frac{1}{\gamma^2}(\bm b+\bm s')^2+(vt-z')^2}.
\end{align}
The center-of-mass of the target nucleus is assumed to move along the $-z$ 
direction with the velocity $\bm v=(0,0,-v)$ and each nucleon of the target 
nucleus is also assumed to follow a straight-line trajectory.  
The above field can be considered  a superposition of fields with various 
frequencies. 

A Fourier analysis of the electric field gives 
\begin{align}
\bm E(\bm r, \omega)|_{\bm r=0}=\frac{1}{2\pi}\int_{-\infty}^{+\infty}dte^{i\omega t}
\bm E(\bm r, t)|_{\bm r=0}.
\end{align}
Performing $t$-integration with Eq.~(\ref{efzero}) leads to 
\begin{align}
&\bm E(\bm r, \omega)|_{\bm r=0}\notag \\
&=-\frac{Z_Te\xi}{\pi  bv}\int d{\bm r}' e^{i\frac{\omega}{v}z'}\rho_T(\bm r')
\left[\hat{\bm p}K_1(p)+i\frac{\hat{\bm v}}{\gamma}K_0(p)
\right],
\end{align}
where $\hat{\bm v}=\bm v/v$ and 
\begin{align}
\bm p=\xi(\hat{\bm b}+\frac{1}{b}\bm s').
\end{align}
The number of equivalent photons modified by the finite charge distribution 
is obtained as 
\begin{align}
&\tilde{N}(\bm b, \omega)\notag \\
&=\frac{Z_T^2\alpha}{\pi^2}\big(\frac{c}{v}\big)^2 \frac{\xi^2}{\omega b^2}\left[{\tilde{\bm K}}_1(\xi,\omega)\cdot {\tilde{\bm K}}_1(\xi,\omega)+\frac{1}{\gamma^2}{\tilde{K}}_0^2(\xi,\omega)\right],
\end{align}
where
\begin{align}
&{\tilde{\bm K}}_1(\xi,\omega)=\int d{\bm r}' e^{i\frac{\omega}{v}z'}\rho_T(\bm r')
\hat{\bm p}K_1(p),\notag \\
&{\tilde{K}}_0(\xi,\omega)=\int d{\bm r}' e^{i\frac{\omega}{v}z'}\rho_T(\bm r')K_0(p).
\label{K.func}
\end{align}
Here $\rho_T(\bm r')$ is assumed to be invariant with respect to the 
reflection of $z' \to -z'$, which guarantees that both 
${\tilde{\bm K}}_1(\xi,\omega)$
and ${\tilde{K}}_0(\xi,\omega)$ are real.
The integration in Eq.~(\ref{K.func}) is easily performed by 
  expanding  $\rho_T(\bm{r})$ in terms of a sum of Gaussians.
See Appendix for details.
The ratio, 
$r(\bm b, \omega)=\tilde{N}(\bm b, \omega)/N(\bm b, \omega)$, gives the 
change of the photon-number spectrum as a function of $b$ and $\omega$:
\begin{align}
r(\bm b, \omega)= \frac{{\tilde{\bm K}}_1(\xi,\omega)\cdot {\tilde{\bm K}}_1(\xi,\omega)+\frac{1}{\gamma^2}{\tilde{K}}_0^2(\xi,\omega)}{K_1^2(\xi)+\frac{1}{\gamma^2}K_0^2(\xi)}.
\label{ratio.EPM}
\end{align}
 At large $\bm{b}$ that exceeds the charge radius of the target,
 we numerically find that
 the ratio has no incident-energy dependence and goes to a constant
independent of $b$. Note that $\int_0^\infty\,d\omega\tilde{N}(\bm{b},\omega)
=\int_0^\infty\,d\omega N(\bm{b},\omega)$ holds at large $\bm{b}$.
We will examine $r(\bm b, \omega)$ in Sec.~\ref{finite.charge.dist}.

\subsubsection{Coulomb breakup reaction probability}

The Coulomb breakup probability of Eq.~(\ref{EPMprob.eq}) is replaced by 
including the finite-charge distribution as follows:
\begin{align}
  P_C(\bm{b})=\sum_{\lambda}\int_0^\infty
  \,d\omega\, r(\bm{b},\omega)
  N_{E \lambda}(\bm{b},\omega)\sigma_{E \lambda}(\omega).
\label{CBprob.eq}
\end{align}
Here we assume that the finite distribution applies equally 
to all the multipoles.
Since the EPM is formulated in a classical way, the
probability $P_C(\bm{b})$ exceeds unity at small $b$.
To avoid this unphysical problem, we multiply the Coulomb breakup probability 
by the survival probability $|e^{i\chi(\bm{b})}|^2$ of the colliding nuclei~\cite{Bertulani93,Bertulani04}
\begin{align}
  \sigma_C=\int d\bm{b}\, P_C(\bm{b})
 |e^{i\chi(\bm{b})}|^2.
\label{Coul.bu.X}
\end{align}
This ansatz is more natural than introducing a sudden cut-off 
impact parameter that is usually taken as a sum of the nuclear radii
of the projectile and target nuclei.

We have discussed the Coulomb excitations of the projectile nucleus by the 
target nucleus. We have to consider the other way around, that is, the 
Coulomb field of the projectile excites the target because 
a measurement excluding such process can not be possible. 
As was done in Ref.~\cite{Horiuchi16}, both 
the Coulomb breakup cross sections of the projectile and target nuclei 
are added incoherently to the nuclear breakup cross section.
$\sigma_{E\lambda}(\omega)$ of the target nucleus is calculated in 
exactly the same manner as that of the projectile nucleus.
It may be likely that the incoherent sum leads to 
some overestimation of $\sigma_C$.
If mutual excitations of both projectile and target nuclei
are considered, it may not be valid to assume
that such excited nuclei generate the same 
photon-number spectrum as the one employed in Eq.~(\ref{photon.eq}).
Instead, they produce somewhat weaker field each other,
leading to the reduced Coulomb breakup cross section.

\section{Results and discussions}
\label{results.sec}

\subsection{Comparison of the EPM with point- and finite-charge distributions}
\label{finite.charge.dist}

\begin{figure}[th]
\begin{center}
\epsfig{file=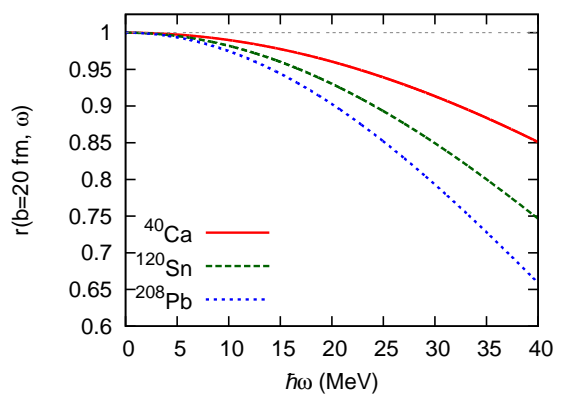,scale=1.2}
\caption{Ratio of the photon numbers $r(b=20\, {\rm fm}, \omega)$ of finite- and point-charge distributions of $^{40}$Ca, $^{120}$Sn, and $^{208}$Pb targets as a function of $\hbar\omega$. See Eq.~(\ref{ratio.EPM}).}
\label{EPMratio.fig}
\end{center}
\end{figure}

To show the effect of the finite-charge distribution
in the EPM, we display in 
Fig.~\ref{EPMratio.fig} the ratio $r(\bm{b},\omega)$ at $b=20$ fm 
as a function of the excitation energy of the projectile nucleus.
For small $\hbar\omega \lesssim 5$ MeV, the 
point- and finite-charge distributions give almost equal photon numbers. 
For $\hbar\omega\simeq 10$-15 MeV where the giant dipole resonance appears,
approximately 5\% reduction is obtained for
$^{208}$Pb target. With increasing $\omega$ further 
suppression occurs for a heavier target nucleus.  
The calculated Coulomb breakup cross sections of $^{120}$Sn 
incident at 100-1000$A$\,MeV are reduced
by 1-3\%, 3-4\%, 4-5\% for $^{40}$Ca, $^{120}$Sn, and $^{208}$Pb targets,
respectively, compared to the case of the point charge.
Hereafter we employ the EPM with the finite-charge distribution. 

\subsection{Systematics of total reaction cross sections}
\label{rcs.sec}

\begin{figure*}[th]
\begin{center}
\epsfig{file=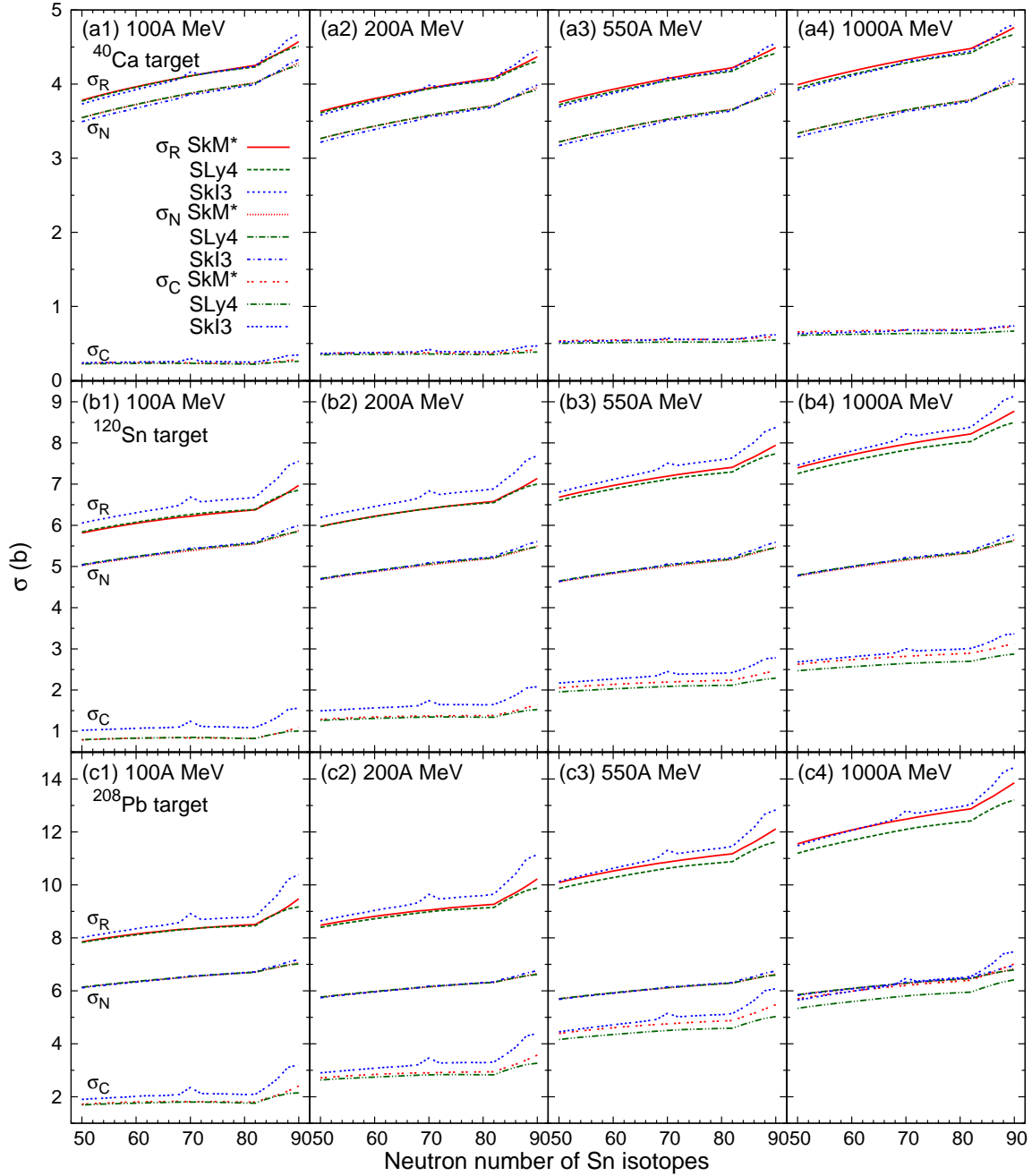,scale=0.9}
\caption{Total reaction ($\sigma_R$), nuclear breakup ($\sigma_N$),
  Coulomb breakup ($\sigma_C$) cross sections
  of Sn isotopes, $^{100-140}$Sn, incident on (a1)-(a4) $^{40}$Ca,
  (b1)-(b4) $^{120}$Sn, and (c1)-(c4) $^{208}$Pb targets at the incident energies
  of 100, 200, 550, and 1000$A$ MeV.
  The SkM*, SLy4, and SkI3 interactions are employed.
}
\label{rcs.fig}
\end{center}
\end{figure*}

Figure~\ref{rcs.fig} displays $\sigma_R$, $\sigma_N$, and $\sigma_C$
of Sn isotopes incident on (a1)-(a4) $^{40}$Ca, (b1)-(b4) $^{120}$Sn,
and (c1)-(c4) $^{208}$Pb targets
at the incident energies of 100, 200, 550, and 1000$A$\,MeV. 
At all incident energies,
the cross sections increase as the neutron number increases.
All the Skyrme interactions give almost the same results for Ca
target because the nuclear breakup contributions dominate.
For Sn and Pb targets, the Coulomb contribution increases and the 
interaction dependence shows up through $\sigma_C$ although 
$\sigma_N$ is insensitive to the interaction.
At the lower-incident energies,
the cross sections calculated with the SkM* and SLy4 interactions
are almost the same, whereas 
those with the SkI3 interaction behave differently from the others.
At the higher-incident energies, the SkM* and SLy4 interactions give 
different cross sections.
The different behavior of $\sigma_R$ with the incident energy 
suggests some change of structure on the Sn isotope chain.

Both the nuclear and Coulomb breakup cross sections increase
gradually as the neutron number increases.
All panels of Fig.~\ref{rcs.fig} show kink behavior at $N=82$
where the neutron $0h_{11/2}$ orbit is fully occupied.
The isotope dependence of $\sigma_N$ is rather moderate, 
reflecting the increase of its matter radius.
The $\sigma_C$ on  $^{40}$Ca target
also shows smooth dependence on the neutron number, but for 
$^{120}$Sn and $^{208}$Pb targets it exhibits a rapid increase at $N>82$.
This behavior corresponds to
the sudden appearance of the low-lying $E1$ strengths~\cite{Ebata14}.

The enhancement of the low-lying $E1$ strength can be understood
by considering the neutron level structure around the Fermi surface.
Though the HF+BCS model mixes the single-particle orbits
near the Fermi surface, we discuss it with
the dominant neutron orbits for the sake of simplicity.
In the mass region of $N=70$-82, the outermost neutrons are filled in
the $0h_{11/2}$ orbit. At $N>82$, the $1f_{7/2}$ orbit
accommodates further neutrons up to $N=90$.
With the SkI3 interaction,
the Fermi energy becomes very small,
accounting for larger enhancement of the low-lying $E1$ strength at $N>82$,
compared to those with the SkM* and SLy4 interactions
(See Ref.\cite{Ebata14} or Fig.~\ref{integrands.fig}
in Sec.~\ref{resp.sec}).

  This excitation mechanism is similar to that found in
  $^{22}$C~\cite{Inakura14} in which the $E1$ strength is governed by
  the single-particle excitations from the outermost $sd$ orbits,
  $1s_{1/2}$ and $0d_{5/2}$, which are energetically almost degenerate.
  The enhancement of the $E1$ strength is found
  as the Fermi energy decreases due to the spatial extension of
  the $sd$ orbits.
  In case of $^{134}$Sn, the root-mean-square (rms) radii
  of the outermost single-particle orbit,
  $1f_{7/2}$, are 5.96, 6.13, and 6.44\,fm
  with the SkM*, SLy4, and SkI3 interactions, respectively.
  The rms radius with the SkI3 interaction extends very much
  compared to the others, accounting for the large enhancement of
  the low-lying $E1$ strength at $N=84$.
  The corresponding single-particle energies are
  $-$3.21, $-$2.15, and $-$1.53\,MeV for SkM*, SLy4, and SkI3 interactions,
  respectively. The rms radius is well correlated
  with the single-particle energy.
  In contrast,  the rms radius of the fully occupied
  $0h_{11/2}$ orbit in $^{134}$Sn remains at almost the same values:
  5.57, 5.61, and 5.67\,fm with
  the SkM*, SLy4, and SkI3 interactions, respectively.
  Since those neutrons are deeply bound at $-$7.8 to $-$8.6\,MeV and
  the radii do not change drastically at $N \leq 82$,
  the interaction dependence of the low-lying $E1$ strength
  is small at $N \leq 82$.

A bump of $\sigma_C$ at $N=70$ appears only with the SkI3 interaction, and it 
is due to an increase of the $E3$ strength function.
Pairing correlations always play a role
of suppressing a sudden structure change
with increasing  neutron number~\cite{Horiuchi16}.
The pairing effect actually vanishes at $N=70$
for the SkI3 interaction,
giving the sudden increase of the $E3$ cross section.

It should be noted that $\sigma_C$ 
becomes very large and comparable to $\sigma_N$ especially 
for large-$Z$ targets at high incident energies
because of the increase of the photon numbers~(\ref{photon.eq}). 
With $^{208}$Pb target, the $\sigma_C$ is almost equal to $\sigma_N$
at incident energies higher than $\sim$500$A$\,MeV.
This suggests that the information of the $E\lambda$
strength function can be observed by measuring $\sigma_R$ 
at different combinations of the incident energy and the target nucleus. 

\subsection{Coulomb multipole excitations}
\label{multi.sec}

\begin{figure*}[th]
\begin{center}
  \epsfig{file=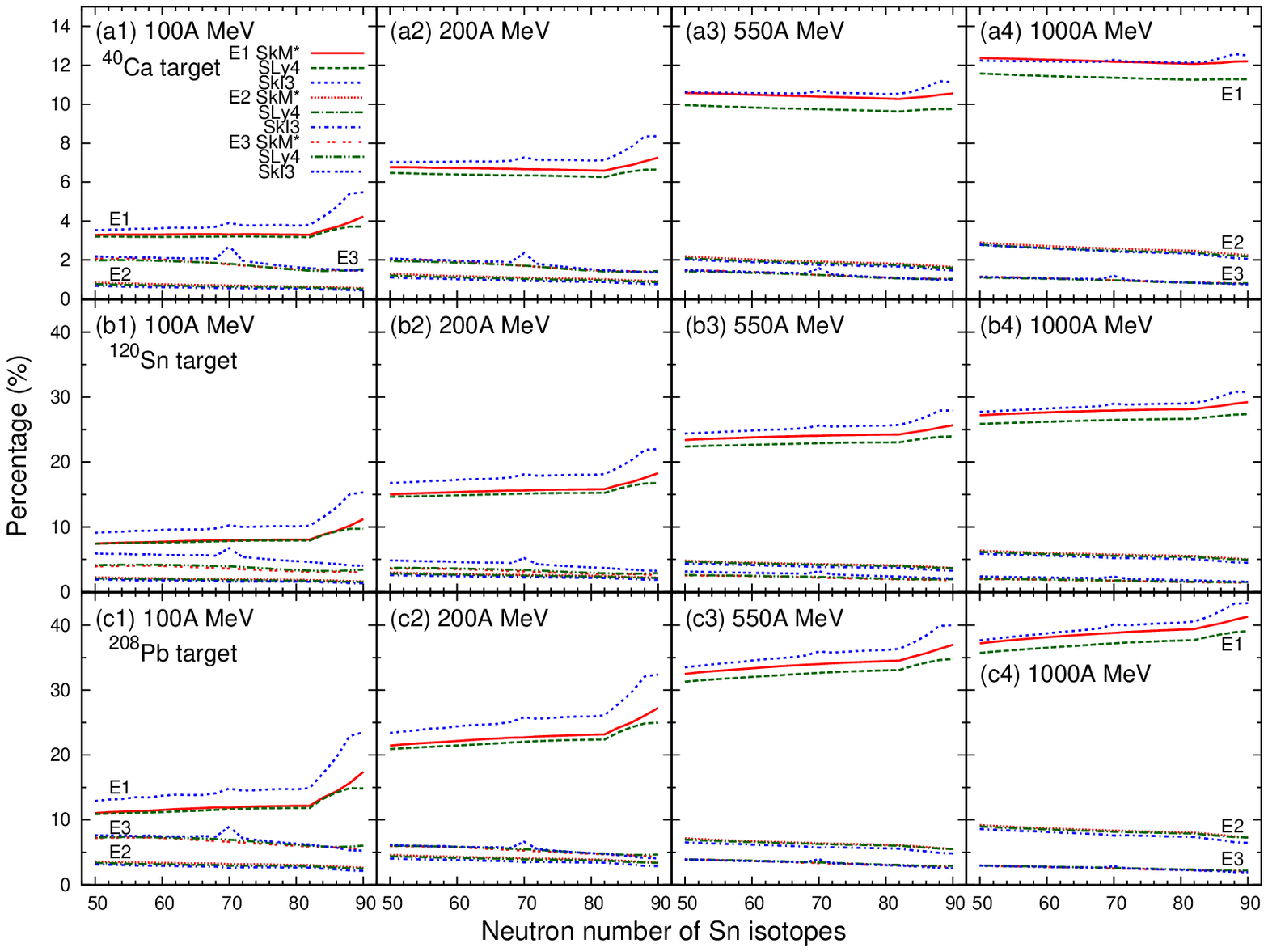,scale=1}
  \caption{Percentages
    of the Coulomb breakup cross sections with
    electric multipoles, $E1, E2$, and $E3$, 
    in the total reaction cross sections of Sn isotopes, $^{100-140}$Sn, incident 
    on (a1)-(a4) $^{40}$Ca, (b1)-(b4) $^{120}$Sn, and (c1)-(c4) $^{208}$Pb targets
    at the incident energies 
    of 100, 200, 550, and 1000$A$\,MeV.
    The SkM*, SLy4, and SkI3 interactions are employed.}
\label{frac.fig}
\end{center}
\end{figure*}

The Coulomb multipole excitations are expected to play an important role
in particular at the low-incident energies~\cite{Bertulani88,Horiuchi16}.
To quantify the contribution of each multipole,
we plot in Fig.~\ref{frac.fig} 
the percentage  of $E\lambda$ Coulomb breakup
cross section compared to the total reaction cross section.
The $E1$ contributions are dominant for all cases although their percentages
depend on the choice of the Skyrme interaction,
target, and incident energy.
At $N=82$,
we see kink behavior which becomes more evident 
as the incident energy is lowered.
The $E2$ and $E3$ percentages show
almost constant behavior and do not so much depend on the neutron number.
The $E3$ contribution is even larger than the $E2$ contribution
when the Sn isotopes are incident on $^{40}$Ca,
  $^{120}$Sn, and $^{208}$Pb targets at the incident energies
  of 100 and 200$A$\,MeV.
This can be explained by the following two factors:
In spherical nuclei, the low-lying $E2$ strengths
are suppressed~\cite{BM2,RS80}. In fact,
all Sn isotopes considered in this paper
have a spherical shape~\cite{Ebata14,Horiuchi16}.
The second is the behavior of the photon-number spectrum
which will be discussed in Sec.~\ref{resp.sec}.
At the high-incident energies, the $E2$ and $E3$ contributions are small, 
approximately one order of magnitude smaller than the $E1$ contribution.
As the incident energy decreases,
the $E2$ and $E3$ contributions compared to the $E1$ become larger.
In the case of $^{208}$Pb target at 100$A$\,MeV, the contribution of 
the higher multipole excitations is comparable to that of $E1$.
Although the isotope dependence of the total reaction cross sections
is dominated by the $E1$ contributions,
the higher multipole contributions have to be included for a 
quantitative evaluation of the cross sections,
especially at the low-incident energy.

To test the validity of our approach, we compare theory with measurement.
Only few experimental data of the total reaction cross section
involving heavy projectile and target nuclei are available in literature. 
The total reaction cross sections of
$^{118}$Sn+$^{40}$Ca and $^{208}$Pb+$^{40}$Ca collisions incident at 77$A$\,MeV 
are tested. The $\sigma_R$ ($\sigma_C$) values calculated with the SkM*, SLy4, 
and SkI3 interactions are, in
units of barn, 4.20 (0.23), 4.20 (0.22), and 4.18 (0.24)
for $^{118}$Sn+$^{40}$Ca, and 5.46 (0.47), 5.43 (0.45), and 5.47 (0.52) 
for $^{208}$Pb+$^{40}$Ca, respectively. The interaction dependence is 
negligibly small. 
The corresponding experimental $\sigma_R$ values 
are 4.89$\pm$0.53 and 5.33$\pm$0.50~\cite{Kox87},  
in fair agreement with the theoretical ones.
The theoretical cross sections may be further improved by including 
higher multipole contributions ($\lambda > 3$) as the incident energy is low.

\subsection{Coulomb breakup and $E\lambda$ strength functions}
\label{resp.sec}

\begin{figure*}[th]
\begin{center}
\epsfig{file=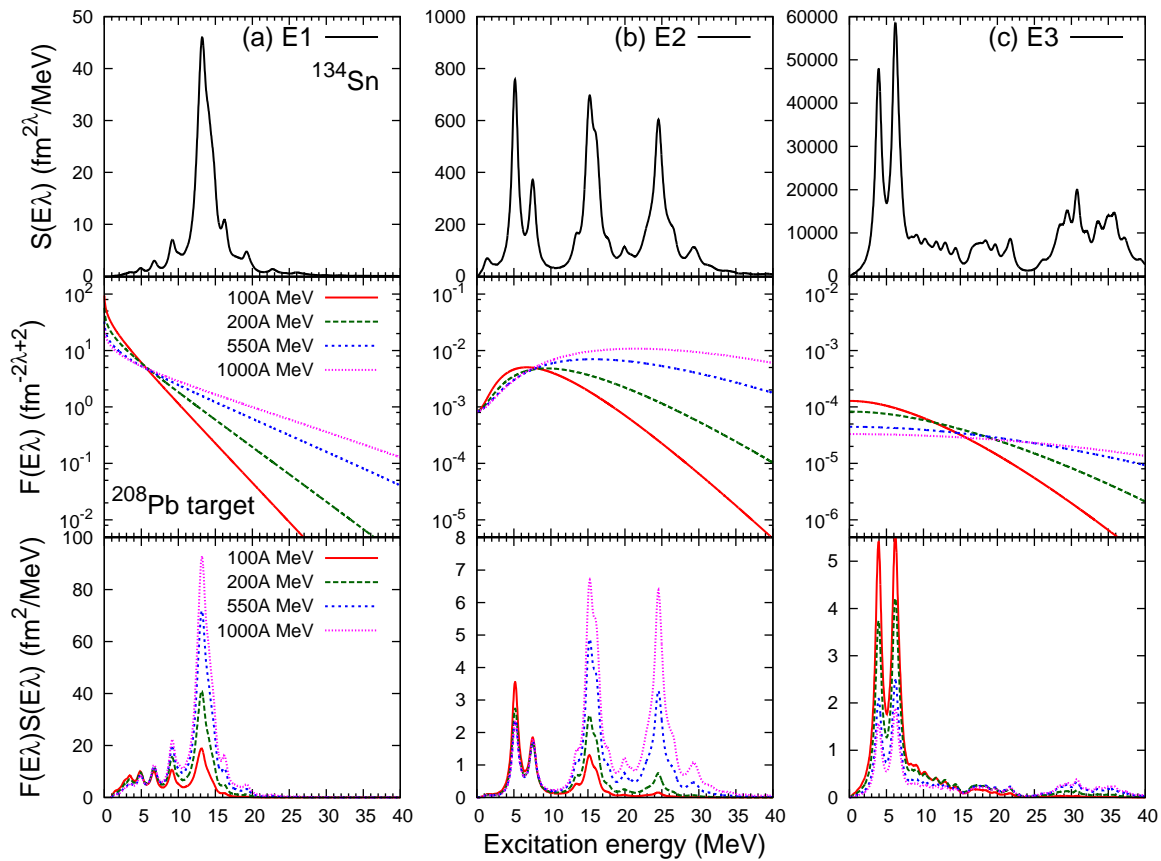,scale=1}
\caption{Contributions of the electric-multipole strengths of $^{134}$Sn 
  to the Coulomb breakup cross section ($\sigma_C$) by $^{208}$Pb target 
  as a function of the excitation energy: (a) $E1$, (b) $E2$, and (c) $E3$.
  See Eqs.~(\ref{sigmaCvsFS}) and (\ref{probe.eq}). The weight function 
  $F(E\lambda)$ is plotted at the incident energies
  of 100, 200, 550, and 1000$A$\,MeV. The SkI3 interaction is employed. 
}
\label{integ.fig}
\end{center}
\end{figure*}

All discussions in the previous subsection
can be understood by making explicit the nuclear structure information 
contained in the Coulomb breakup cross section. 
For this purpose, we rewrite $\sigma_C$~(\ref{Coul.bu.X}) 
as an integral of the $E\lambda$ strength function, $S(E\lambda; \omega)$, 
over the excitation (photon) energy:
\begin{align}
  \sigma_C=\sum_{\lambda}
  \int_0^\infty d\omega\,F(E\lambda;\omega)S(E\lambda;\omega),
\label{sigmaCvsFS}
\end{align}
where the weight function $F(E\lambda; \omega)$ contains the dynamical 
aspect of the Coulomb breakup reaction, especially the equivalent photon 
numbers:  
\begin{align}
  F(E\lambda; \omega)&=\frac{(2\pi)^3 (\lambda+1)}{\lambda[(2\lambda+1)!!]^2}\omega^{2\lambda-1}\notag \\
&\times  \int\,d\bm{b}\,r(\bm{b},\omega)N_{E\lambda}(\bm{b},\omega)
   |e^{i\chi(\bm{b})}|^2.
  \label{probe.eq}
\end{align}
The expression of $\sigma_C$ as an integral over $\omega$ is 
more natural than that over ${\bm b}$. This is because the Coulomb 
breakup occurs even at large impact parameter due to its long-range force
and rather we are interested in the nuclear response as a function of the 
excitation energy.

Figure~\ref{integ.fig} plots 
$S(E\lambda)$, $F(E\lambda)$, and their product $F(E\lambda)S(E\lambda)$,
$E\lambda$ Coulomb breakup cross section per unit energy,
as a function of $\omega$ for the Coulomb breakup of $^{134}$Sn ($N=84$) by 
$^{208}$Pb target.  The SkI3 interaction is employed.
For the sake of simplicity, only the excitation 
of $^{134}$Sn is taken into account, whereas the contribution of
the $^{208}$Pb excitation is ignored.
The $S(E1)$ exhibits the so-called pygmy dipole resonance below $10$ MeV
and the giant dipole resonance peak at around 13 MeV.
The $E1$ weight function $F(E1)$
decreases rapidly as the excitation energy increases.
As the incident energy increases, the falloff of $F(E1)$ with the 
excitation energy becomes more gentle because of the increase of 
the photon number. The $E1$ Coulomb breakup cross section per unit energy,
$F(E1)S(E1)$, does not depend on the
incident energies at low excitation energies up to about 7 MeV, 
while it is enhanced with the increasing incident energy
in the giant dipole resonance region.
The excitation-energy dependence of the 
$E1$ Coulomb breakup cross section
at 1000$A$\,MeV is similar to that of $S(E1)$.
The Coulomb breakup cross section at the high-incident energy
can therefore be a probe of the non-energy weighted $E1$ sumrule, 
which is closely related to the radii of the 
proton distribution~\cite{BM,Lipparini89}.

We turn to the $E2$ contribution.
$S(E2)$ shows some low-lying peaks at about 5 MeV
and two large peaks at the higher energy region, while 
$F(E2)$ has almost no vital dependence on the 
incident energy up to 10 MeV but becomes larger and larger beyond 10 MeV 
as the incident energy increases. Thus the $E2$ cross section 
increases with the increase of the incident energy. 
Since the $E2$ strengths are suppressed in spherical Sn isotopes
and therefore $F(E2)S(E2)$ is small,
the $E2$ contribution to $\sigma_C$ is much smaller than $E1$.

In the $E3$ case, $F(E3)S(E3)$ at the low excitation energy 
becomes smaller and smaller with increasing incident energy, in  
contrast to the $E1$ and $E2$ cases. This is understood from 
the excitation-energy dependence of $F(E3)$. 
This specific energy dependence plays a role in enhancing the 
$E3$ contribution   
at the low-incident energy as displayed in Fig.~\ref{frac.fig}.

\begin{figure*}[th]
\begin{center}
\epsfig{file=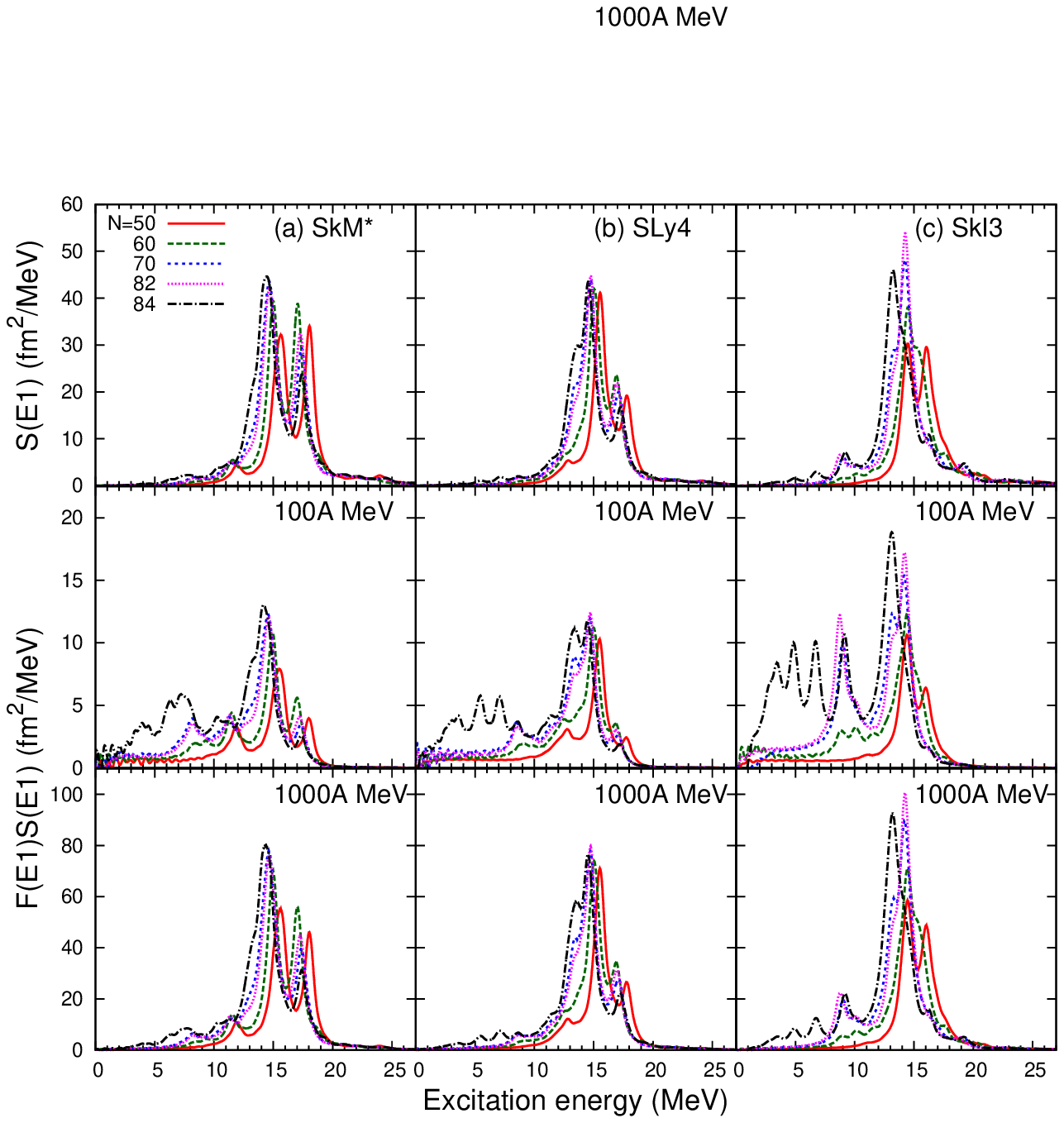,scale=1}
\caption{Comparison of the electric-dipole ($E1$) contributions of Sn isotopes, 
  $^{100,110,120,132,134}$Sn, to the Coulomb breakup cross sections by 
  $^{208}$Pb target. The $E1$ strength functions $S(E1)$ are plotted 
  as a function of the excitation energy for three Skyrme interactions, 
  (a) SkM*, (b) SLy4, and (c) SkI3, and for each case two incident energies of 
  100$A$ and 1000$A$\,MeV are chosen to draw the Coulomb breakup 
  cross sections per unit energy $F(E1)S(E1)$. 
}
\label{integrands.fig}
\end{center}
\end{figure*}

Since the $E1$ contributions dominate in $\sigma_C$,
it is interesting to examine the extent to which $S(E1)$
and $F(E1)S(E1)$ change with the neutron number. 
Figure~\ref{integrands.fig} displays the results of 
$^{100,110,120,132,134}$Sn calculated with the three Skyrme interactions. 
The low-lying $E1$ strength contributes significantly to the reaction
probability at the low-incident energy  
because the photon numbers or $F(E1;\omega)$
concentrate at the low-energy region.
At 100$A$\,MeV the low-lying 
strength is significantly enhanced
compared to that in the giant dipole resonance
region, while at 1000$A$\,MeV no such enhancement occurs and 
the reaction probability distribution
is similar to that of the $E1$ strength distribution.
Since the photon number in the low-excitation energy region
becomes large at the low-incident energy, 
the information on the low-lying $E1$ strength can possibly be
obtained by a measurement involving the $E1$ Coulomb breakup 
process at the low-incident energy.

  \subsection{Total reaction cross sections of Ca and Ni isotopes}
\label{rcsCaNi.sec}

\begin{figure*}[th]
\begin{center}
\epsfig{file=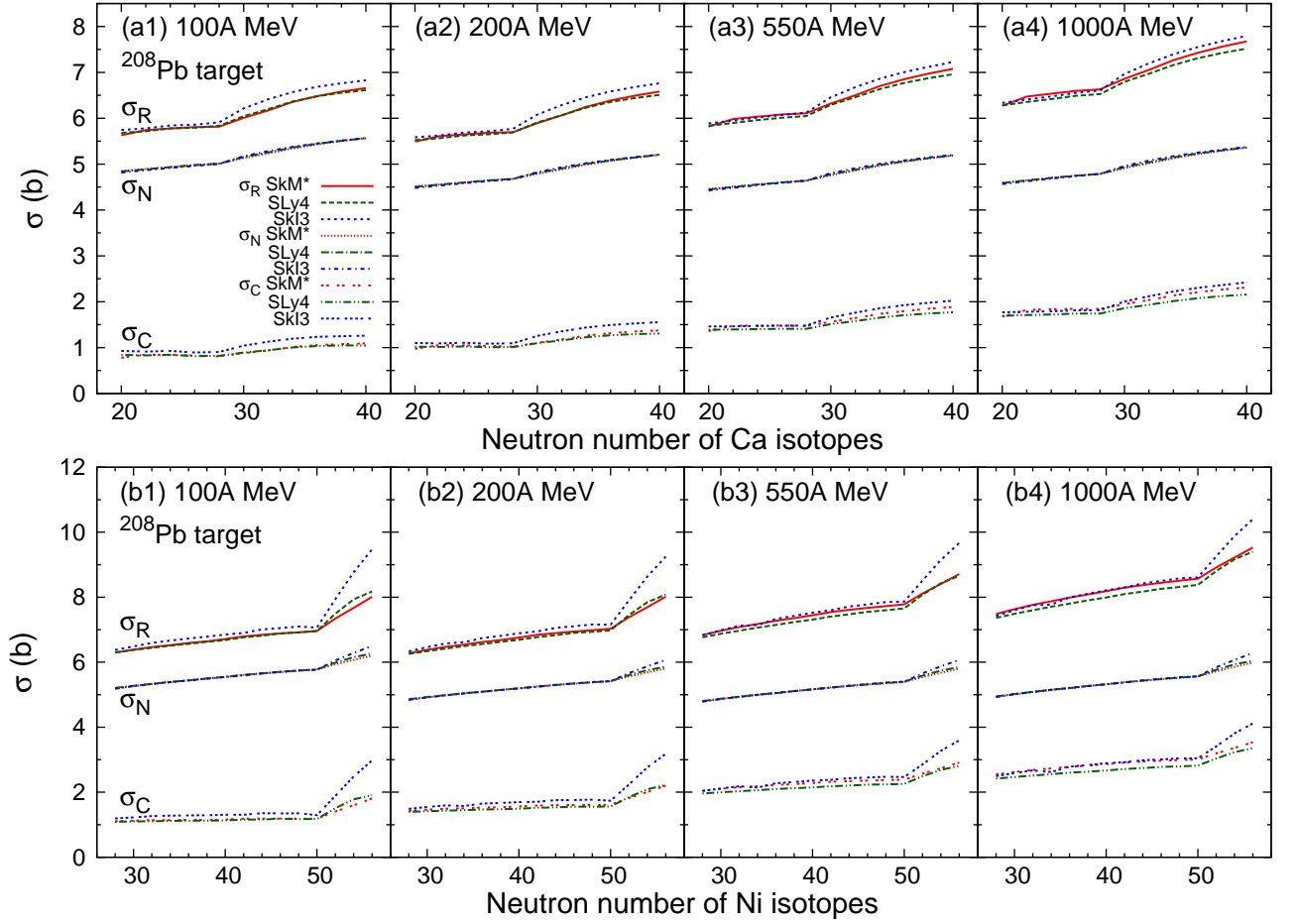,scale=1}
\caption{Same as Fig.~\ref{rcs.fig} but for (a1)-(a4) $^{40-60}$Ca
  and (b1)-(b4) $^{56-84}$Ni incident on $^{208}$Pb target.
}
\label{rcsCaNi.fig}
\end{center}
\end{figure*}

The mechanism of the cross section enhancement in the Ca and Ni isotopes
is similar to that of the Sn isotopes
but different single-particle orbits are involved.
Figure~\ref{rcsCaNi.fig} plots
the total reaction cross sections of Ca and Ni isotopes
incident on $^{208}$Pb target with various incident energies.
Though the Coulomb breakup cross sections are not as large as
those of Sn isotopes because of smaller-$Z$ values of Ca and Ni isotopes,
as expected, large enhancement of the Coulomb breakup cross sections
is found at $N> 28$ and $N > 50$ for Ca and Ni isotopes, respectively.
The enhancement becomes more prominent with lowering the incident energy.
At $N=28$ ($N=50$), $0f_{7/2}$ ($0g_{9/2}$) orbit is fully occupied
and the weakly-bound neutron orbits in the higher major shell
play a primarily important role at $N>28$ ($N>50$).
Similarly to the Sn case, the enhancement is due to sudden changes
of the Fermi energies or rms radii of the outermost neutron orbit.
For the most prominent case, $^{80}$Ni ($N=52$),
the single-particle energies and rms radii of the dominant outermost orbit,
$1d_{5/2}$, are $-$2.72, $-$1.62, $-$1.31\,MeV, and 5.46, 5.71, and 6.05\,fm
for the SkM*, SLy4, and SkI3 interactions, respectively.
In fact, the SkI3 interaction gives a drastic increase at $N=52$.
For $^{50}$Ca ($N=30$), those of the outermost $1p_{3/2}$ orbit
are $-$5.76, $-$6.59, and $-$5.16\,MeV, and 4.60, 4.47, and 4.79\,fm
for the SkM*, SLy4, and SkI3 interactions, respectively.
Since the single-particle energy (radius) of the outermost neutron orbit
in $^{50}$Ca is not as small (large) as that of $^{80}$Ni,
the $E1$ transition is suppressed.
Therefore, the enhancement of the Coulomb breakup cross section
at $N> 28$ of the Ca isotopes
is not so significant compared to that at $N> 50$ of the Ni isotopes.

\section{Conclusions}
\label{conclusions.sec}

The low-lying $E1$ strength crucially depends on the
shell structure near the Fermi surface.
We have investigated the extent to which 
information on the $E1$ strength function
of neutron-rich Ca, Ni, and Sn isotopes
is imprinted on the total reaction cross sections.
The nuclear breakup contributions are calculated based on the Glauber model
with density distributions obtained by the Skyrme-Hartree-Fock+BCS method.
The Coulomb multipole excitations of $E1$, $E2$, and $E3$ are
also included with the use of the equivalent photon method (EPM), 
where the point-charge is replaced
by the realistic finite-charge distribution and  
the strength function corresponding to the $E\lambda$ excitation is obtained by 
the canonical-basis-time-dependent-Hartree-Fock-Bogoliubov method.

No significant dependence of the $E1$ strength distribution
appears with the small-$Z$ target, $^{40}$Ca,
because the nuclear breakup cross section,
which mostly reflects the matter radius,
dominates in the total reaction cross section.

We have found that the 
low-lying $E1$ strength gives relatively large contribution 
to the total reaction cross section at the low-incident energy.
At low-incident energy,
the contributions of higher multipoles also get larger.
In such a case, however,
the neutron number dependence of the total reaction cross sections
is still governed by the low-lying $E1$ strength because
the $E2$ and $E3$ contributions are not strongly dependent on
the number of the neutrons, showing almost constant behavior.
As the incident energy increases, the contribution 
from the strength in the giant dipole resonance region becomes large.
The multipole excitations higher than $E1$ get small
with increasing incident energy.
With use of Pb target,
the nuclear and Coulomb contributions of Sn isotopes
become comparable and the $E1$ contribution is dominant
at the incident energy higher than 500$A$\,MeV

With use of the larger-$Z$ targets, Sn and Pb,
the difference of the Skyrme interaction or
the shell structure near the Fermi surface can be seen clearly in
the Coulomb breakup cross sections, which strongly depend
on the excitation mechanism of the projectile and target nuclei.
A comparison of the theory and experiment
is desired to understand the shell structure
of Ca, Ni and Sn isotopes beyond $N=28$, 50 and 82, respectively.
Since the Coulomb breakup cross sections strongly depend on
the low-lying $E1$ strength or the interaction employed
at $N>28$, 50, and 82 in the Ca, Ni, and Sn isotopes, respectively,
it also gives strong constraint on the effective interaction.

In the present paper, we have discussed only spherical nuclei in which
the $E2$ transitions are suppressed and change moderately 
on the neutron number. 
If the projectile nuclei exhibit different deformation,
the $E2$ contribution may become large and changes significantly 
as a function of the neutron number.
Further investigation for such systems is an interesting
subject for future.

Our calculation of the Coulomb breakup cross section is 
performed on the basis of the EPM.
Since the mutual Coulomb excitation of the projectile and
target nuclei are treated independently in the present paper,
its validity has to be tested by a comparison with experiment.
Though our calculations agree with the few existing data,
more data on accurate total reaction cross sections
of nucleus-nucleus collisions are needed.  
As discussed in Ref.~\cite{Brady16}, the channel coupling effects
becomes important in the Coulomb breakup process at the low-incident energy. 
It is certainly desirable to develop a consistent theory that 
can describe nucleus-nucleus inclusive Coulomb excitations.

\acknowledgments

The work was in part supported 
by JSPS KAKENHI Grant Number JP15K05072.

\appendix
\section{Evaluation of Eq.~(\ref{K.func})}

The aim of this appendix is to carry out the integration in 
Eq.~(\ref{K.func}) for a finite-charge distribution.
For $\rho_T(\bm r)$ given as a superposition of Gaussians
\begin{align}
\rho_T(\bm r)=\sum_i C_i e^{-a_i r^2},
\end{align}
the integration~(\ref{K.func}) is reduced to 
the following form:
\begin{align}
&\int d{\bm r} \, e^{i\frac{\omega}{v}z}e^{-ar^2}K_0(|\xi(\hat{\bm b}+\frac{1}{b}\bm s)|)\notag \\
&\ =2\pi \sqrt{\frac{\pi}{a}}\frac{b^2}{\xi^2}
\exp\big(-\frac{\omega^2}{4av^2}-ab^2\big)\notag \\
&\times \int_0^{\infty}dp\, p\,
e^{-a\frac{b^2}{\xi^2}p^2}I_0\big(2a\frac{b^2}{\xi}p\big)K_0(p),
\label{int.K0}
\end{align}
and
\begin{align}
&\int d{\bm r} \, e^{i\frac{\omega}{v}z}e^{-ar^2}\widehat{\xi(\hat{\bm b}+\frac{1}{b}\bm s)}K_1(|\xi(\hat{\bm b}+\frac{1}{b}\bm s)|)\notag \\
&\ =2\pi \sqrt{\frac{\pi}{a}}\frac{b^2}{\xi^2}
\exp\big(-\frac{\omega^2}{4av^2}-ab^2\big)\notag \\
&\ \times \hat{\bm b}\int_0^{\infty}dp\, p\,
e^{-a\frac{b^2}{\xi^2}p^2}I_1\big(2a\frac{b^2}{\xi}p\big)K_1(p),
\label{int.K1}
\end{align}
where $I_m$ is the $m$-th order modified Bessel function 
of the first kind. The $p$-integration in Eqs.~(\ref{int.K0}) 
and (\ref{int.K1}) can easily be done numerically.

\end{document}